\documentstyle[12pt]{article}

\begin{document}

\author{V. V. Okorokov}
\title{On the incompatibility of experiments \\ 
confirming certain conclusions \\ 
of the general relativity. }
\date{}
\maketitle

\begin{center}
Instutute of Theoretical and Experimental Physics,\\
B. Cheremushkinskaya 25, Moscow, 117259, Russia\\
E-mail: okorokov@vxitep.itep.ru
\end{center}

\begin{abstract}
\noindent
Qualitativ arguments are presented which show the incompatibility of
the positive results obtaned in experiments on the gravitational
redshift of photones and in experiments investigating the behavior of
clocks in the gravitational field.
\end{abstract}

The present note originated when the author had been preparing
proposals of experiments aimed at using in fundamental research the
effects of coherent exitation of fast atoms or nuclei passing through
the crystal \cite{1} - \cite{4}. In particular, the interest was in
verification of the equivalence between the gravitational field and
accelerated frame at the values of acceleration
$10^{20}\div 10^{21}$cm/sec$^{2}$. Trying to analyse possible results
of such experiments and their relation to the wellknown experiments
on certain predictions of the General Relativity (G.R.) the author
encountered and interesting and somewhat paradoxical situation.

Unfortunately discussions which were initiated by the author and which
lasted for quite a long time did not resulted in a reasonable
clarification of the paradoxical situation. Hence the author considers
it necessary to focus on the attention of scientific media on the
question to be explained in what follows.

As it is wellknown experiments now considered as classical \cite{5} -
\cite{10} have confirmed the predictions of the G.R. on the frequency
shift of photons moving along the gradient of the gravitational
potential and on the difference of the frequencies of the clocks placed
at the points with different gravitational potential. The gravitational
shift of the photon frequency $\Delta \nu /\nu =g\,H/c^2$ has been
measured in the known experiments of Pound and Rebka \cite{5}, \cite{6}
and Vessot and Levine \cite{7} - see Fig.~1.

The frequency shift $\Delta \nu $ of photons emitted (by $Fe^{57}$
nuclei \cite{5}, \cite{6} or by hydrogen atoms in the hydrogen
frequency standard (HFS) \cite{7}) during the elevation (or descend) to
the altitude H in the gravitational field was detected by comparison
with the reference frequency of the `generator' placed at the point
where the photon was detected. In \cite{5} the role of the `generator'
of the reference frequency placed at the altitude $H$ was played
by $Fe^{57}$ nuclei (identical to the nuclei-emitters with the energy
14 keV at $H=0$), while in \cite{7} it was played by HFS-emitter of
photons placed at the earth level. In both experiments it was
considered as obvious that the level spacing in nuclei \cite{5},
\cite{6} or atom \cite{7} does not depend upon the gravitational
potential (G.P.).

We remind that the stability of the frequency of the emitted photons
and the accuracy of the determination of the frequency of the detected
photons is undoubtedly defined by the stability of level spacing in
nuclei \cite{5}, \cite{6} or atoms \cite{7}.

Without this implicit assumption on the independence of the level
spacing on the G.P. the interpretation (and even the performance) of
these experiments is impossible since the small shift of the photon
frequency can be detected only comparing it with
the \underline{constant} reference frequency which is determined by
the constant level spacing of nuclei or atoms.

It seems that three alternative interpretations of the experimental
results on the gravitational frequency shift of photons are feasible:

\begin{description}
\item[a).]  when moving upwards the photon frequency changes in line
with the equation of G.R. $\Delta \nu /\nu =g\,H/c^2$,
while the level spacing of nuclei and atoms does not depend on G.P.

\item[b).]  the photon frequency remains unchanged, while the levels
of nuclei and atoms follow the G.P. according
to $\Delta \nu /\nu =g\,H/c^2$;

\item[c).]  both the photon frequency and the nuclear levels are
changed (in these case different options are possible depending upon
the sign and the value of the above changes when moving in the
gravitational field).
\end{description}

As one can infer from \cite{5}, \cite{6}, \cite{7}, these paper tacitly
and with no doubts imply the option a).

The dependence of the clock rates on the G.P. (quantitatively it is
given by the twin-equation $\Delta T /T =g\,H/c^2$) was investigated
in experiments \cite{8}, \cite{9}, \cite{10} by the comparison of the
counts of the two high-accuracy frequency standards at the point with
certain fixed value of the G.P. and subsequent elevation of one of the
frequency standards for a certain period of time to the point with
different value of the G.P. (to the altitude H of several kilometers).

The difference in the run of the two devices after they were returned
to the same point (Fig.1) quantatively confirms the dependence of the
clock rate on G.P. in line with the G.R. prediction.

The interpretation of these experiments is directly related to the fact
that the positions of the levels (which determine the rates of the
frequency standards) \underline{depend on the value of the G.P.} at the
point where the atoms are placed. The atoms play here the role of
`clocks' which measure how the time runs at different altitudes in the
gravitational field of the earth.

Thus the interpretation of the Pound and Rebka \cite{5} and Vessot and
Levine \cite{7} experiments is based on the `seemingly evident'
assumptions (the invariance of the nuclear and atomic levels with the
G.P. variations) which are in sharp contrast with the known G.R. result
on the different clock rate at the points with different G.P. This G.R.
result was also experimentaly confirmed in \cite{8}, \cite{9}, \cite{10}
where it was shown that the \underline{levels} of at least
\underline{atoms change with G.P.} (the atoms are in fact clocks which
react to the change of the G.P.!).

Thus the positiv experimental results on the gravitational photon
frequency shifts \cite{5}, \cite{6}, \cite{7} on one hand, and
experimental results on the gravitational change in the clock rate
\cite{8}, \cite{9}, \cite{10} on other hand, are unfortunately
incompatible.

If the atomic and nuclear levels do not depend on the G.P. - then
experiments \cite{5}, \cite{6}, \cite{7} must yield positive result,
while \cite{8}, \cite{9}, \cite{10} - negative. If on the contrary the
positions of the atomic and nuclear levels do depend on the G.P., then
experiments \cite{5}, \cite{6}, \cite{7} can not lead to positive
result to which lead \cite{8}, \cite{9}, \cite{10}. Getting
simultaneously positive results in
experiments \cite{5}, \cite{6}, \cite{7} and \cite{8}, \cite{9},
\cite{10} is impossible since the positions of the atomic and nuclear
levels can not at the same time be dependent and independent
upon the G.P.

The paradoxical and problematic physical situation which has emerged
practicaly `from nothing' results in the whole chain of important
physical consequences which are yet precocious to discuss. Still one
point worth reminding - it is just the movement in the effective
gravitational field due to acceleration (the equivalence principle!)
which is at the core of the twin-paradox [11].

The above incompatibility between the results of experiments which are
already considered as being classical calls for the necessity of
additional experimental confirmation using alternative methods.
This role may be played by the experiments on the coherent exitation
of the levels of the fast atoms and relativistic nuclei in
crystals \cite{1} - \cite{4}. In such experiments the projective nuclei
serves as a clock and its rate is compared to that of the atoms in
crystal by means of subsequential interactions.

The sharp resonance form of the interaction enables to single out from
the level shift of the moving atoms the component which is due to the
changing G.P. which is in turn caused by the effective deacceleration
of the nuclei inside the media.

The work was fulfilled with the support of Russian
Foundation for Basic Research. Grant No. 98-02-16782.


\newpage

\begin{thebibliography}{99}
\bibitem{1}  V.V. Okorokov, Yad. Fiz. 2(1965) 1009[Sov.J.Nucl.Phys.
2(1966)716].

\bibitem{2}  V.V. Okorokov, Zh. Eksp. Teor. Fiz. Pis'ma Red. 2(1965)175
[JETP Lett. 2(1965)111]. 

\bibitem{3}  C. Moak, S. Datz, O.H. Crawford, H.F. Krause, P.F. Dittner, J.
Gomez del Compo, J.A. Biggerstaff, P.D. Miller, P. Hvelplund and H.
Knudse. Phys. Rev.{\bf A.} 19 (1979) 843.

\bibitem{4}  Yu.L. Pivovarov, A.A. Shirokov, Yad. Fiz. {\bf 44}(1986)882;
Sov. J. Nucl. Phys. {\bf 44} (1986) 569.

\bibitem{5}  R.V. Pound, G.A. Rebka, Phys. Rev. Lett. 4 (1960) 337.

\bibitem{6}  R.V. Pound, J.L. Snider, Phys. Rev. {\bf B.} 140 (1965) 788.

\bibitem{7}  R.F.C. Vessot, M.N. Levine, Gen. Rel. Gravit. 10 (1979) 181.

\bibitem{8}  C.O. Alley, {\em at all.} Experimental Gravitation. Proc. conf.
at Pavia, (sept. 1976), ed. B. Bertotti, Academic Press, 1977.

\bibitem{9}  J.C. Hafele, R.E. Keating, Science 177 (1972) 166-168.

\bibitem{10}  V.B. Braginskiy, A.G. Polnarev, Udivitelnaya gravitatsiya,
Moscow., Nauka, Gl. Red. Fiz. Mat. literatury, 1986, 71.

\bibitem{11}  A. Einstein, Naturwiss., 1918, 6, 697-702.
\end{thebibliography}
\end{document}